# Hydrophobic hydration driven self-assembly of Curcumin in water: Similarities to nucleation and growth under large metastability, and an analysis of water dynamics at heterogeneous surfaces

**Milan Kumar Hazra, Susmita Roy and Biman Bagchi[*]**

Solid State and Structural Chemistry Unit, Indian Institute of Science, Bangalore-560012, India
Address for Corrospondence:profbiman@gmail.com

## *Abstract*

**As the beneficial effects of curcumin have often been reported to be limited to its small concentrations, we have undertaken a study to find the aggregation properties of curcumin in water by varying the number of monomers. Our molecular dynamics simulation results show that the equilibrated structure is always an aggregated state with remarkable structural rearrangements as we vary the number of curcumin monomers from 4 to 16 monomers. We find that the curcumin monomers form clusters in a very definite pattern where they tend to aggregate both in parallel and anti-parallel orientation of the phenyl rings, often seen in the formation of β-sheet in proteins. A considerable enhancement in the population of parallel alignments is observed with increasing the system size from 12 to 16 curcumin monomers. Due to the prevalence of such parallel alignment for large system size, a more closely packed cluster is formed with maximum number of hydrophobic contacts. We also follow the pathway of cluster growth, in particular the transition from the initial segregated to the final aggregated state. We find the existence of a metastable structural intermediate involving a number of intermediate-sized clusters dispersed in the solution. We have constructed a free energy landscape of aggregation where the metatsable state has been identified. The course of aggregation bears similarity to nucleation and growth in highly metastable state. The final aggregated form remains stable with the total exclusion of water from its sequestered hydrophobic core. We also investigate water structure near the cluster surface along with their orientation. We find that water molecules form a distorted tetrahedral geometry in the 1$^{st}$ solvation layer of the cluster, interacting rather strongly with the hydrophilic groups at the surface of the curcumin. The dynamics of such quasi-bound water molecules near the surface of curcumin cluster is considerably slower than the bulk signifying a restricted motion as often found in protein hydration layer.**

## I. INTRODUCTION

Curcumin, the main ingredient of Indian curry spice, is widely known for its therapeutic activity. It has been used since antiquity to treat various fatal diseases along with minor sickness like inflammation on skin. Several preclinical studies proved that curcumin may become a potent drug ingredient in several fields of medicinal biology such as cancer, Alzheimer's disease, ratinal degeneration etc[1, 3-14]. The molecular origin of this diverse activity is still not completely understood.

The molecular formulae of curcumin is $C_{21}H_{20}O_6$ [IUPAC ID: (1E, 6E)-1,7-bis(4-hydroxy-3-methoxyphenyl)-1,6-heptadiene-3,5-dione]. Curcumin has several tautomeric forms, including a 1, 3-diketo form and two equivalent Enol forms. The Enol and Keto forms are shown in **Fig. 1**. Both in the solid and solution phase, Enol form is more stable energetically. Due to the presence of a large number of aromatic rings, curcumin is sparingly soluble in water. Experimental investigation of aggregation of curcumin with spectroscopic evidence is also known. Hydrophobic nature of the phenyl rings of curcumin along with hydrogen bond formation ability of the side groups are proposed to be the reasons for the aggregation in water [14]. Several studies on exited state photo physics of curcumin in organic solvents reveal that the fluorescence lifetime of curcumin and deuterated curcumin in methanol has a decay component of 130 ps which is attributed to the solvation of the species and exited state intra molecular hydrogen atom transfer. Fluorescence spectra of curcumin show a red Stokes shift which may be due to the formation of aggregates [16].

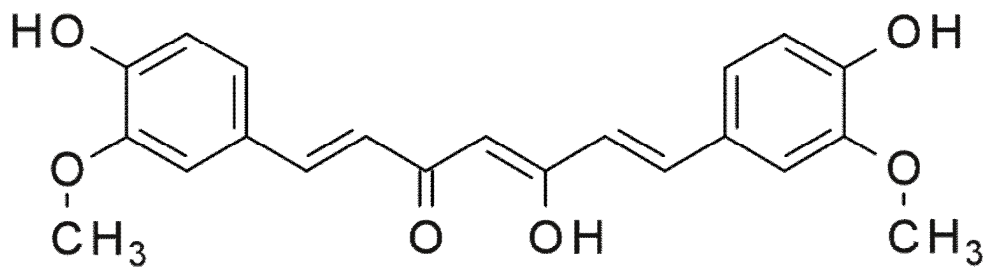

(a)

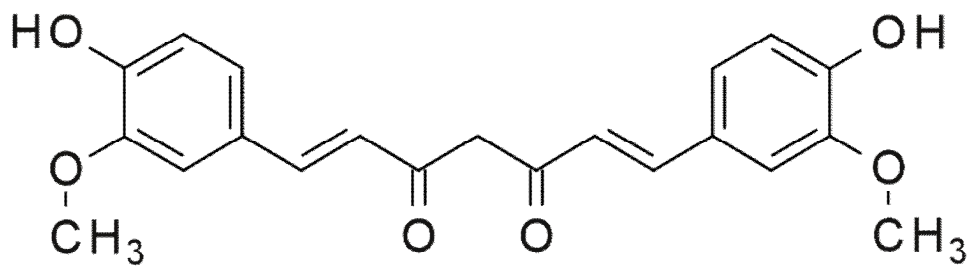

(b)

**FIG. 1. Molecular structure of curcumin. It can exist in nature in two tautomeric forms: (a) the enol form and (b) the keto form.**

A number of studies revealed that dose dependent response to curcumin is limited to its low level of concentration [10,11]. In the lower dose limit it seem to function efficiently by associating with some proteins (eg. beta amyloid) preventing their hydrophobic aggregation that causes Alzheimer's and Parkinson diseases[12-14]. A survey of the literature recommends the daily doses over a 3-month period of up to 12 grams as a safe intake limit of curcumin. Several reports also clearly state that high dose limit is often toxic and undesirable[10]. Although the direct molecular reason for the usefulness of low dose is not clear, we hypothesize that competitive energetics and patterns involved in the self association of curcumin and its association with other proteins, at high concentration, is the plausible responsive factors that crucially determine the detrimental high dose effect.

It is rather difficult to comment on and compare the relative binding energies involved in curcumin-curcumin and curcumin-protein associations as there is no reliable reports available providing the accurate information about the favorable binding sites of curcumin. It should be emphasized that the hydrophobic association of curcumin-curcumin or curcumin-protein is promoted by the surrounding water molecules. The present study indeed shows that curcumin monomers form clusters through the association of phenyl rings[17, 18]. In fact, curcumin-curcumin association proceeds through such orientation of phenyl rings that allows maximum hydrophobic contacts. The strong propensity towards the association of curcumin molecules for each other favors the formation of significantly large cluster which, in turn, makes individual curcumin molecule unavailable for protein binding and for other biological functions or therapeutic activities. Although the hydrophobic association between curcumin monomer is expected, the formation of hydrophobic core with hydrophilic groups projected outside is not a priory obvious. On the other hand, this is pointing towards a general pattern of other hydrophobic aggregation mediated through phenyl or phenyl alanine type of aromatic rings.

Due to the formation of hydrophobic core, curcumin association behaves a bit like a rather long chain protein, rich in hydrophobic residues. We observe a marked slow-down of interfacial water dynamics at the surface curcumin cluster than that of bulk water. Such similar restricted movement is also obtained for "biological water" on the protein surface [1, 2, 19-21].

## II. STRUCTURAL DETAILS OF CURCUMIN AND SIMULATION SETUP

### A. Charge calculation

As the force field parameters for this molecule are not known, we evaluated the charge distribution of the curcumin molecule by using the Gaussian 09 software. The molecule was first optimized in vacuo. Then the optimized structure was reoptimized in a polarization continuum model solvent medium. All the calculations have been done using the B3LYP method with 6-31+g(d,p) basis set. These calculated charges are shown in **Table I**. They were used in the molecular dynamics simulation of the present system. Two different forms of curcumin (namely Keto and Enol) were optimized.

As we have mentioned before that curcumin can exist in two predominant chemical conformations, namely, keto and enol form. The calculated charges and the assigned charge on each respective atom are shown in the supplementary material (see **Fig. S1 and S2** in the Supplementary Material [22]), for both keto and enol forms of curcumin in polarized continuum model of solvent.

**Table I. Calculated values of optimized energy, dipole moment and HOMO-LUMO energy gap for different forms of curcumin in water as well as gas phase.**

| Form | Medium | Energy($10^{-19}$J) | Dipole Moment(D) | Energy difference between HOMO and LUMO($10^{-19}$J) |
|---|---|---|---|---|
| Anti di-keto | Gas Phase | -55103.46 | 0.8 | 5.896 |
| Anti di-keto | Aqueous Soln | -55104.78 | 1.4 | 5.639 |
| Enol | Gas Phase | -55103.942 | 7.6 | 5.209 |
| Enol | Aqueous Soln | -55105.11 | 10.6 | 4.919 |

B. **Molecular dynamics simulation details**

All the simulations were done at 300K temperature and 1bar pressure. The extended simple point charge model was employed to study the water -water interaction [23]. Full atomistic details were retained for every molecule except the hydrogen atoms bound to carbon atoms of the methoxy group. They were treated as united atoms within the Gromos 53a6 force field [24, 25]. To perform the molecular dynamics simulation we have chosen Groningen machine for Chemical Simulation which is a highly efficient and scalable molecular dynamics simulation engine.

We have prepared systems with nearly same mole fraction of curcumin in water but of different size in cubic boxes. After the energy minimization by the method of steepest descent the NPT system was equilibrated for 1 ns. The temperature was kept constant at 300K by using Noose-Hoover thermostat and pressure was kept at 1bar using Parinello–Rahman barostat [26-30]. A production run was carried out by using same NPT system for 10ns. All the properties were extracted from the final trajectory extracted from production run.

All the above simulations were carried out using 2 fs time step. Periodic boundary conditions were applied. All non bonded force calculation employed a grid system for neighbor searching. To perform the calculation of electrostatic interaction Particle Mesh Ewald (PME) was used [31].

## III. HYDROPHOBIC FORCE DRIVEN CURCUMIN AGGREGATION

Experimental investigations over biomolecular aggregations do not provide a great deal of microscopic details that a theoretical model study often does. The present molecular dynamics simulation study not only seeks to understand the hydrophobic interaction rendering the curcumin aggregation, it also focuses on the hydrophobic hydration that offers an inevitable contribution to the stability of that hydrophobic association. Therefore we present different aspects of hydrophobic aggregation first, followed by the effects of hydrophobic hydration.

### A. Hydrophobic core formation and preferred mutual orientations of phenyl rings of curcumin

In **Fig. 2** we have shown the radial distribution function among the curcumin molecules. It is evident from the figure that the monomers are forming an extended hydrophobic core with hydrophilic groups pointing outwards. The hydrophobic core absorbs additional monomers as we increase the system size, and this process seems to continue *ad infinitum.* The figure also shows two prominent maxima at two different positions that indicate different layers of cluster formation distributed in a wide range of the system. To gain further insight into how individual monomers adhere to each other in the aggregate, we calculate angle distribution of two vectors lying on each phenyl plane of two different curcumin molecules. We find a bimodal distribution that provides important information about the structural alignment of two phenyl rings next to each other. From the distribution and also from visual analysis, two types of phenyl-phenyl orientations have been characterized, namely parallel and anti-parallel, depending on the alignment of the two vectors lying on phenyl ring (snapshots in **Fig. 4** clearly shows the alignment.) Populations of such phenyl alignments are shown in **Fig. 3**.

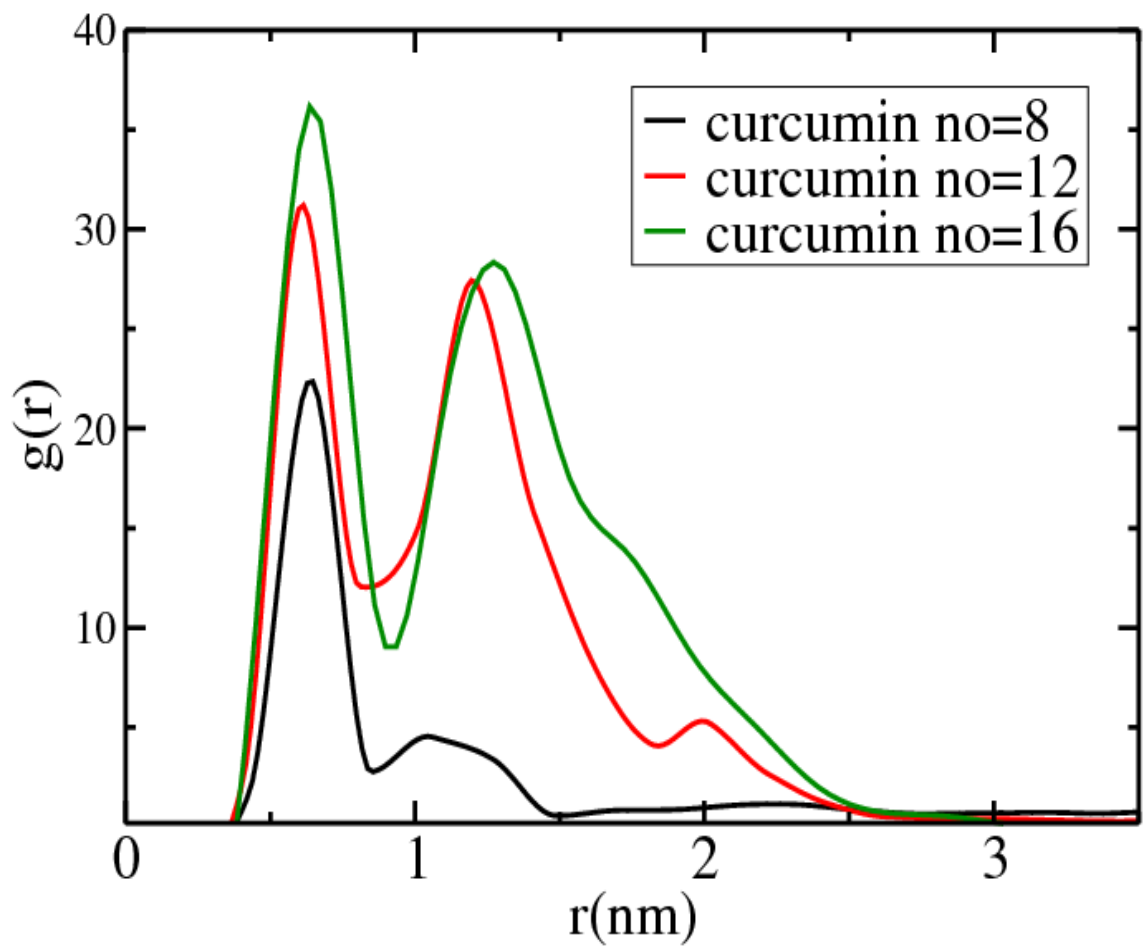

**FIG. 2. Radial distribution function (rdf) between two distinct curcumin molecules in the aggregate, for different cluster sizes. The plot clearly shows aggregation on a nanometer scale. The 1$^{st}$ peak appears at 0.64 nm for 8 curcumin monomers, 0.60 nm for 12 curcumin momomers and 0.65 nm for 16 curcumin monomers. 1$^{st}$ minimum appears at 0.84 nm for 8 curcumin monomers, 0.84 nm for 12 curcumin monomers and 0.916nm for 16 curcumin monomers. With increase in system size two prominent maxima appear, this signifies the emergence of layers in curcumin cluster.**

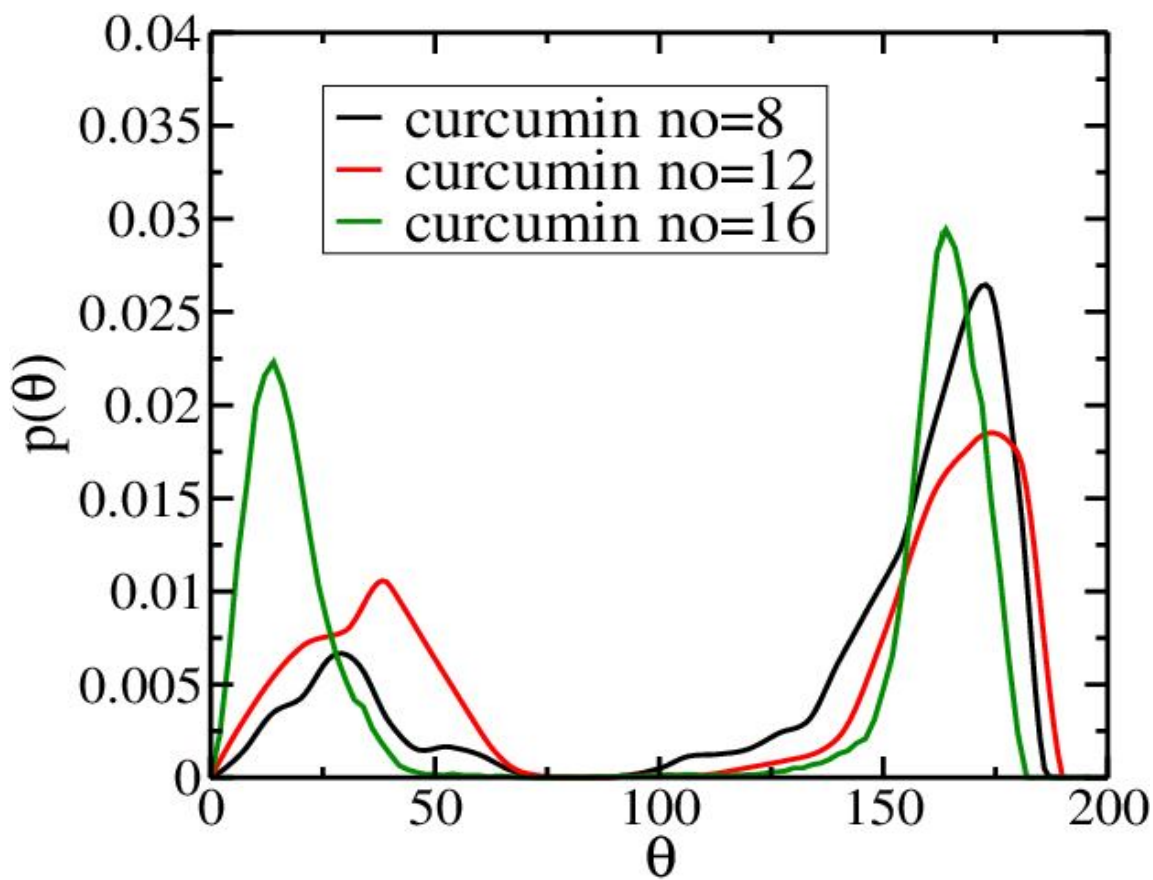

**FIG. 3. Distribution of angles between phenyl-phenyl rings of two different but neighboring curcumin molecules. The angle θ is calculated between two vectors lying on the plane of two different phenyl rings of two distinct curcumin molecules within the cut off distance of 1$^{st}$ minima of RDF of curcumin molecules, according to their various system sizes. Bimodality of the distribution shows two different modes of stacking (discussed in detail below).**

Here we interestingly observe a considerable enhancement in the population of parallel alignments increasing the system size from 12 to 16 curcumin monomers. However, it is important to mention here that although the population of anti-parallel arrangements of phenyl rings slightly increases from 8 to 12, we do not find any significant change in the population of anti-parallel orientation moving from 12 to 16. To address the emergence of parallel orientation predominance more clearly, we have extracted representative snapshots from the equilibrium trajectories of 12 and 16-mer system sizes. We have denoted the parallel phenyl stacking by black circle and anti-parallel phenyl stacking by magenta circle. These snaps (see **Fig. 4**) provide a noticeable prevalence of parallel stacking over the anti-parallel alignment (among the curcumin monomers) in the 16-mer curcumin system. This clearly suggests that as we increase the system size, the phenyl stacking orientation shifts to include more number of phenyl rings by parallel orientation. It is also clear that the formation of such aggregation is mediated by the strong hydrophobic interactions among the phenyl rings of curcumin and the preferred orientation between two rings provides an additional stability for better stability of the cluster than the anti-parallel packing.

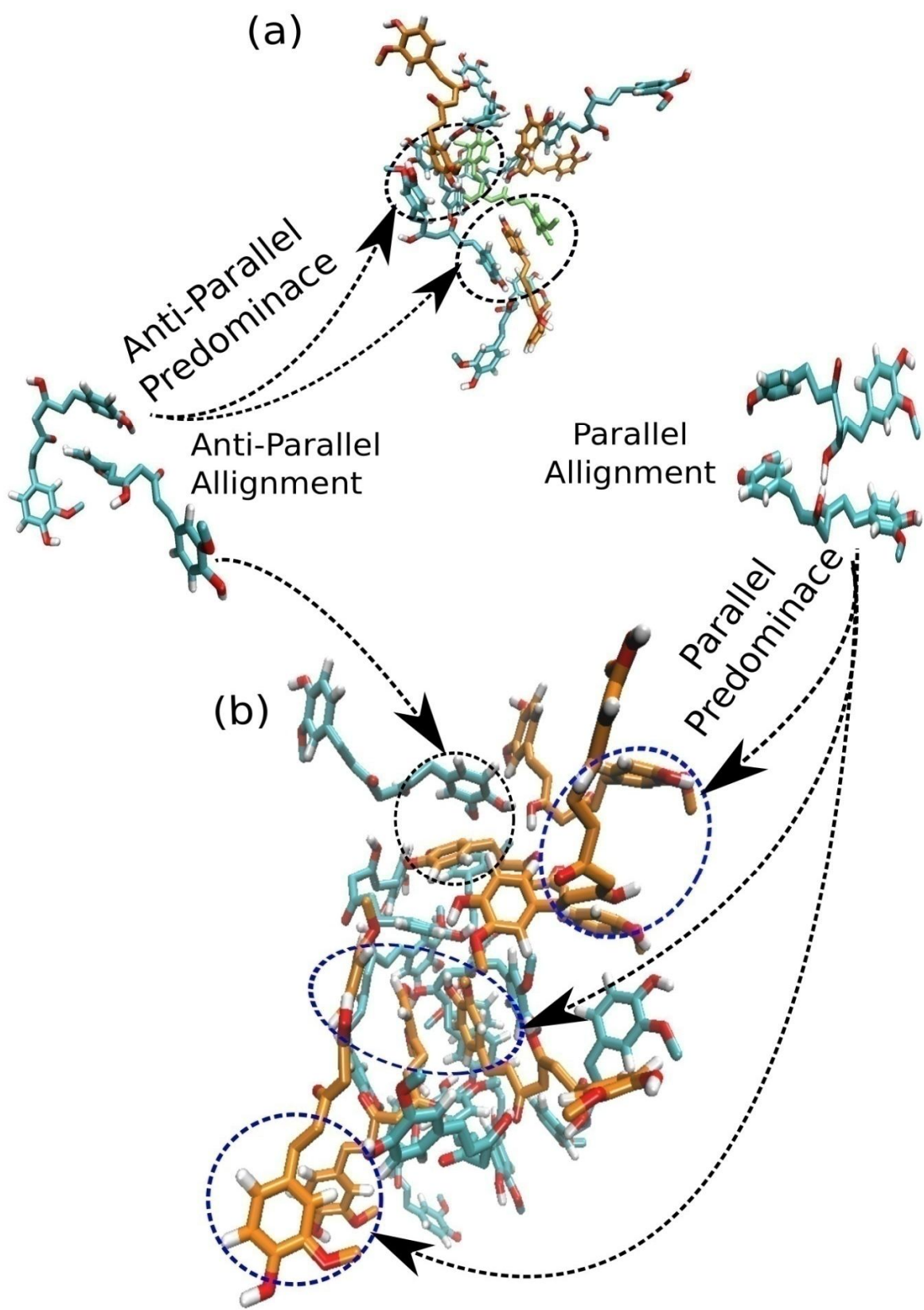

**FIG. 4.** Representative snapshots extracted from simulation trajectories. Parallel and anti-parallel orientations of phenyl rings of two different curcumin molecules in the aggregated state are shown by arrows in (a) Aggregation of 12 curcumin monomers in aqueous solution (b) Aggregation of 16 curcumin monomers in aqueous solution. With small system size only anti-parallel orientation is predominant. However, with growth of cluster size, along with the anti-parallel orientation, parallel oriented curcumin population also begins to emerge in the cluster. Phenyl rings in the surface of a growing cluster acts as the anchor of further addition of curcumin molecules or clusters. Water molecules are not shown here for the clarity of the snapshots.

### B. Cluster growth analysis: from segregation to aggregation

Early experimental and simulation studies on hydrophobicity largely focused on the emergence of stable aggregated structures of several hydrophobic solute like methane, ethane and amphiphilic solutes like phenyl alanine, caffine, to name a few[32-35]. However the course of transition from segregated structures to an aggregated one perhaps has never been followed up adequately. In the present study we track a number of simulation trajectories of 16 curcumin monomers to analyze the sequence of events that leads to the formation of final aggregated structure. In order to understand the molecular interaction among the monomers we computed total number of cluster and the largest cluster size at each snapshot and follow their time evolution (see **Fig. 5 (a)**). As expected that total number of cluster and the largest cluster size progress in a dynamically anti-correlated fashion. *The progression of cucumin aggregation has a fair resemblance with the nucleation phenomena of a liquid where the monomer assemble and dissemble course of action continue until it reaches to a critical cluster size*[36-38]. Although the pathway formation of a stable cluster here is a multistep process, it broadly involves three distinct stages. In the initial stage, the monomers rapidly form a number of segregated clusters dispersed throughout the system. In the next stage, few segregated clusters of comparable sizes enter into a competition among themselves to capture the remaining isolated monomers. Such isolated monomers seem to fail to attach to any of intermediate sized segregated structures (may be due to entropic reason) to congregate with them and remain frustrated for a long time. This kind of segregated assemblies form a metastable (here with lifetime ~10-30ns) configuration where both the number of cluster throughout the system and the size of largest cluster remains largely unaffected or constant. We believe that such configurational intermediates gain both entropic and enthalpic stabilization. In the final or last stage of aggregation, one segregated cluster win over the other to assemble all the remaining fractionated parts. The qualitative free energy landscape of the pathway of cluster growth has been shown in **Fig. 5(b).** The free energy diagram provides a clear evidence of appearance of the configurational intermediate. It presents a barrier separated minima in the free energy landscape, similar to the situation observed in nucleation at large metastability[36-39].

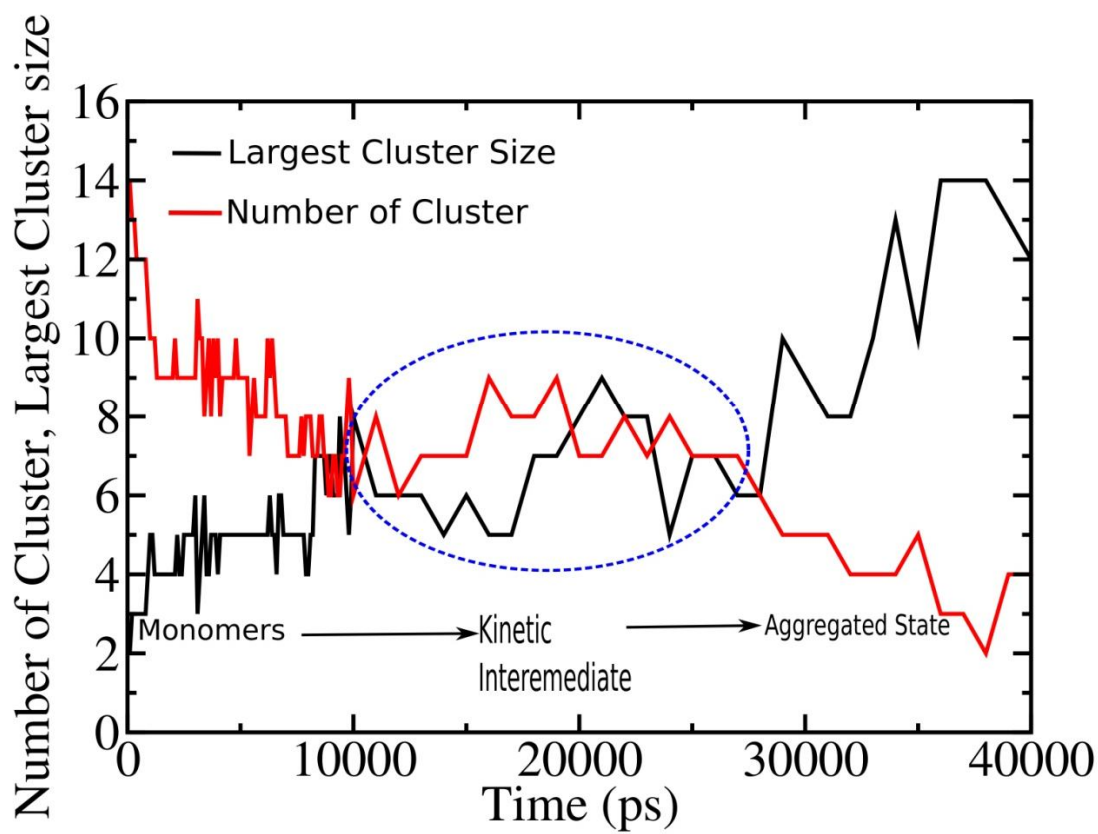

**(a)**

**FIG. 5 (a) :** Multistage dynamics of the temporal evolution of the total number of clusters and the largest cluster size of 16-mer curcumin (N=16) in the course of their aggregation. The total number of clusters and the largest cluster size evolve in a dynamically anti-correlated fashion. The cluster growth pathway goes through a metastable, relatively long-lived intermediate state (shown in the figure by a dashed ellipse) before arriving at the final stable aggregated state where it stays during the rest of the simulation.

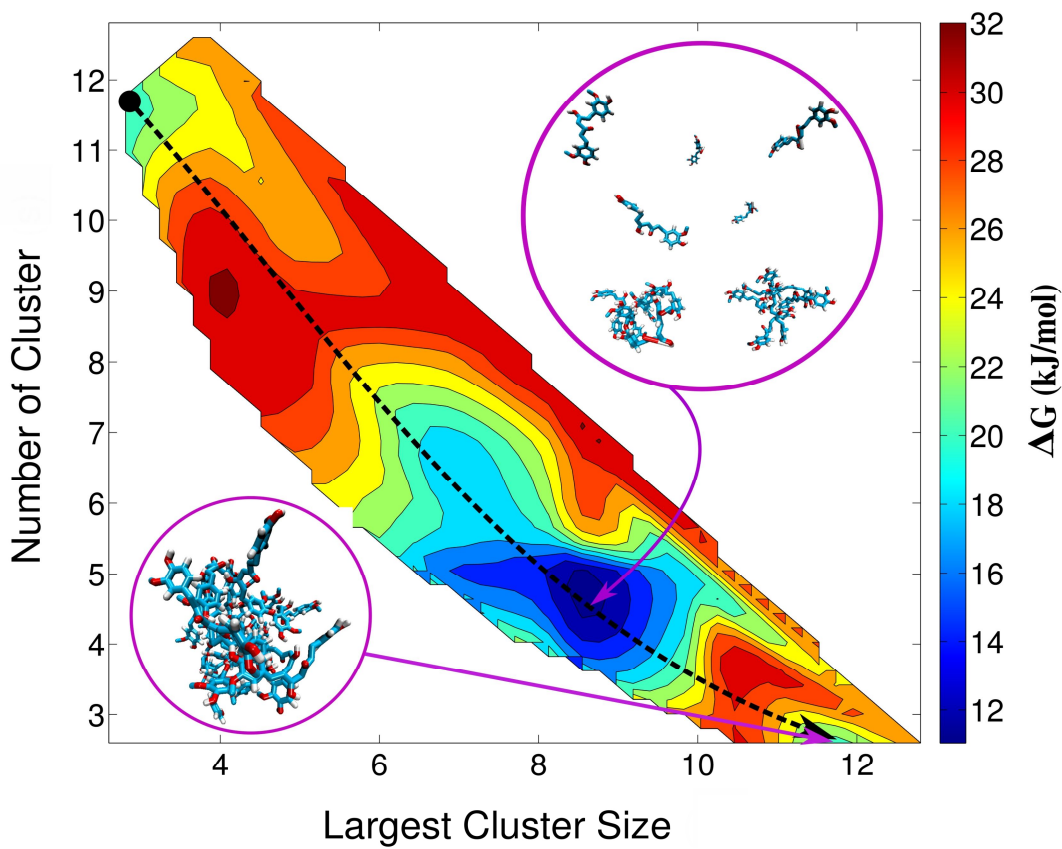

**(b)**

**FIG. 5(b):** Free energy change, ΔG, plotted against relevant order parameters along the cluster growth pathway for N=16, showing configurational degrees of freedom visited. The X-axis and Y-axis is represented as largest cluster size and the total number of clusters respectively at their explicit time. The most visited configurational spaces allow us to derive a probable pathway that might be the lower energy path of curcumin aggregation. The intermediate is demonstrated as it appeared at midstage of the cluster growth to trigger the final aggregation process. Note that due to less population of final aggregated cluster here we are unable to show the depth of the free energy minima.

## C. Clustering as a phase transformation under large metastability

The presence of the transient state observed in **Fig. 5(b)** bears a striking similarity to nucleation of a stable phase from the unstable or metastable phase under large super-saturation (or large metastability)[36-39]. In the latter case also a certain number of intermediate sized clusters forms and serves as the launching point towards the formation of the large liquid-cluster. In the case of nucleation of liquid from gas and solid from liquid, one finds an apparent free energy minimum at intermediate sized cluster.

The reason for the apparent free energy minimum along the cluster growth pathway is probably both kinetic and thermodynamic in origin. As the system is homogeneous initially, the short time growth of a cluster occurs by a random adherence of particles that are close to each other. This random growth process comes to a halt when concentration of monomers surrounding a cluster is fully depleted. Then the system enters a stagnant phase. This bears similarity to the initial phase of Ostwald ripening. Subsequent growth can happen in two ways : (i) The intermediate sized clusters aggregate as in nucleation [36-38], or (ii) disappearance of smaller clusters by evaporation of monomers and deposition of the same on larger clusters. This is the late stage of Ostwald ripening.

In the present case, it appears that the primary growth mode is the first process, as in nucleation, by combination of intermediate sized clusters. A theoretical model of the process was described by several people, notably by Kitahara, Metiu and Ross [39].

## IV. HYDRATION OF CLUSTER

As we have mentioned before, hydration can greatly affect the stability of a cluster [29, 30]. Here we analyze the local water structure surrounding the aggregated curcumin. Due to the presence of some hydrophilic groups pointing outwards, the cluster surface is largely hydrophilic. Local water structure specific to hydrophobic and hydrophilic surfaces are pretty different.

**Fig. 6** shows density of the solvent surrounding the largest cluster relative to the bulk solvent density. It is important to note here that the local structure of interfacial water around the curcumin cluster is quite similar to the arrangement of hydration water surrounding a protein molecule[1,2,19-21]. Here we indeed find that cluster surface is not dewetted [29]. The drying takes place only near the hydrophobic sites that forms an extended core. This is because the large scale of hydrophobicity breaks the hydrogen bonded network. Hence density decreases [40-48]. While analyzing the radial distribution function of curcumin molecule and water we need to consider the fact that, near the surface of hydrophilic units, the immediate hydrophobic neighbor may affect the surrounding water density.

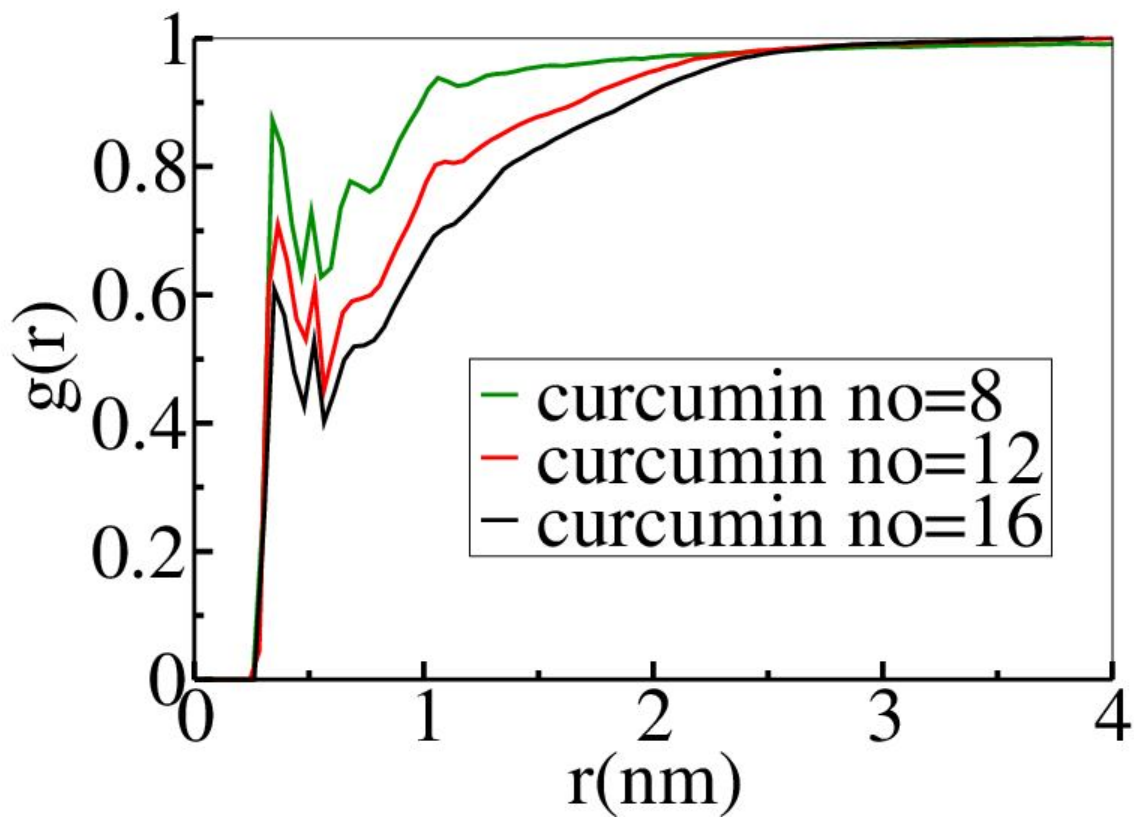

**FIG. 6. Site-site radial distributions function (rdf) of water molecules around the curcumin cluster. The lowering of peak value along with system size indicates that the overall density of surrounding water molecules decreases as the curcumin cluster size increases. The water molecules escape from those curcumin monomers that are participating in hydrophobic core formation. This causes an effective dewetting in the core. Note that, 1$^{st}$ minimum of rdf appears at 0.468 nm for 8 curcumin monomers, 0.478nm for 12 curcumin monomers, 0.468 nm for 16 curcumin monomers.**

A. **Structure of the hydration shell**

In the present case we find that the interfacial water structure is affected by the combined effect of hydrophilic and hydrophobic effect due, to their contiguous presence. In order to quantify microscopic arrangement of the water molecules surrounding a cluster we monitor the extent of tetrahedrality in terms of tetrahedral order parameter [49,50]. The mathematical expression of the tetrahedral order parameter is given below:

$$t_h(i,t) = 1 - \frac{3}{8}\sum_{j=1}^{3}\sum_{k=j+1}^{4}\left(\cos\theta_{jk}(i,t) + \frac{1}{3}\right)^2 \quad (1)$$

where $\cos\theta_{jk}(i,t)$ is the angle of i$^{th}$ water with its two nearest neighbors j and k at a time frame t. Angle distribution of each water with its four nearest neighbors is shown in **Fig. 7.** It determines the extent of deviation of water structure from the ideal tetrahedral network. The equation is defined such that $t_h(i,t)$ can vary -3 to 1. For a completely random orientation six angles will not have any correlation among themselves and the tetrahedrality parameter is zero. Thus, the average tetrahedral order parameter can vary within the range of 0 to 1. The distribution shows an enhancement of three coordinated water in the 1$^{st}$ solvation shell of the solute.

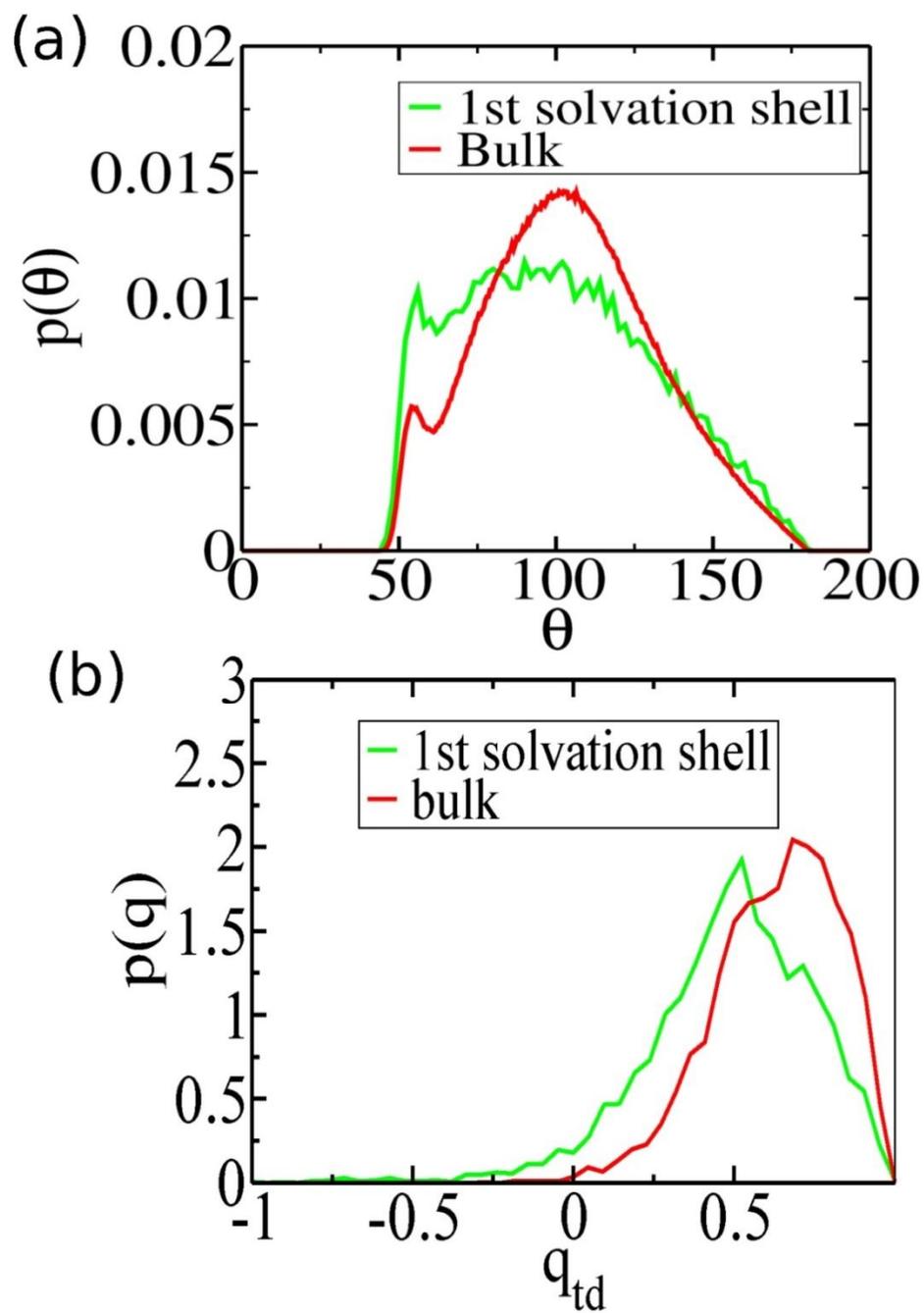

**FIG. 7. (a) O-O-O angle distribution of the water molecules around the entire curcumin cluster surface to determine their degree of tetrahedrality. Note that the enhanced peak ~ 57°  indicates that**

**maximum water molecules in the first solvation shell (shown in green) of curcumin attain the coordination number of less than 4. (b) Local tetrahedral order parameter around the hydrophilic surface of curcumin. Shift of distribution of local tetrahedral order parameter towards lower value is a signature of distorted tetrahedral arrangement. It is important to mention here that in the calculation of tetrahedral order, we predominantly obtain the effect surface hydrophilicity as the cluster surface is largely hydrophilic, with very few hydrophobic residues peeping outward.**

Local tetrahedral order parameter distribution of water near hydrophobic surface confirms the evidence of three coordinated water present in the $1^{st}$ solvation shell. Computation of average tetrahedral order parameter in the $1^{st}$ solvation shell of solute gives a value of 0.49 which is far below the bulk value of 0.60. Evaluation of coordination number for both bulk water and $1^{st}$ solvation shell reveals the fact that in first solvation shell most waters are three coordinated where bulk water molecules achieve four coordinations (see **Fig. S3** in the supplementary material [22]).

### B.  Orientation of water molecules surrounding a cluster

In bulk water each oxygen molecule of water is, on the average, tetrahedrally surrounded by four other oxygen molecules of water through hydrogen bonded network. But in the presence of a large hydrophobic solute or surface, this arrangement of water molecules gets altered drastically. Near a small hydrophobic solute, the interfacial water molecules try to retain their tetrahedral network by minimal distortion or re-rearrangement of the bulk structure.

**Fig. 8** shows the distribution of the angle of orientation of water dipole moments with respect to the normal to the cluster surface. Bimodal distribution with a sharp rise at around $\cos\theta = 0.76$ and a broad distribution over obtuse angle of orientation bear the signature of a clathrate-like hydration shell structure where water molecules are pointing three of their hydrogen bonding sites toward the surface and one site pointing away[51-53]. There is a local cage like hydrogen bonded structure with distinct loss in entropy.

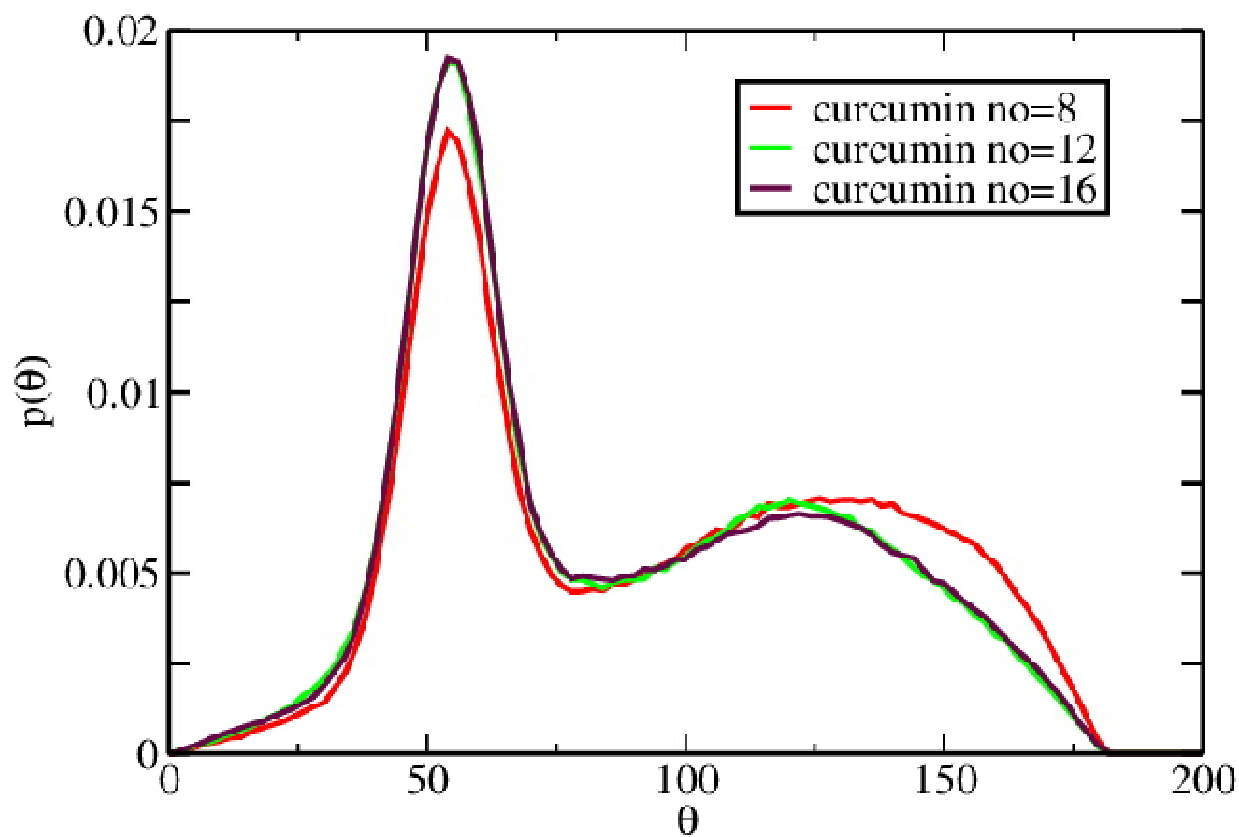

**FIG. 8. Orientational distribution of water molecules in the hydration shell, relative to the normal of the cluster surface. θ is the angle between water dipole and surface normal of the cluster. In the sharp acute angle region, water dipoles are towards the surface whereas broad obtuse part shows water dipoles are towards the bulk. This is similar to the clathrate-like hydration of water found**

**near bio-molecules where three of hydrogen bond forming vectors are towards the surface and one pointing outward.**

### C. Dynamics of interfacial water around curcumin cluster

Whether water is mediating the assembly or not is a critical question necessary to answer. In this section we present the orientational dynamics of water near the surface. We report that the orientational dynamics of water near the surface of the curcumin cluster is slower than that of bulk. **Fig. 9** shows the orientational correlation function of surface waters with respect to bulk. It is interesting to note that though the water molecules deviate from their tetrahedral arrangements, the hydration layer dynamics remains quite slow. This can be explained by observing the fact that solute forms a surface which is largely hydrophilic, with only a few of hydrophobic entities are present. So water maintains its tetrahedral hydrogen bonded network by making one hydrogen bond to hydrophilic surface of curcumin cluster and other three hydrogen bonds to waters surrounding nearest to it. Such surface bound water molecules become highly enthalpically stable and their dynamics becomes quite slow with respect to the bulk. This phenomena is quite similar to what one finds near a protein surface [1,2,39,51,55,56]. We also find that the water density near surface increased almost 1.25 times that of bulk. This is partly driving the arrangement of the solute molecules in the assembly such that water can bind to it if there is any chance of it, preventing segregation of curcumin. It seems to build a cavity of hydrophobic character inside water.

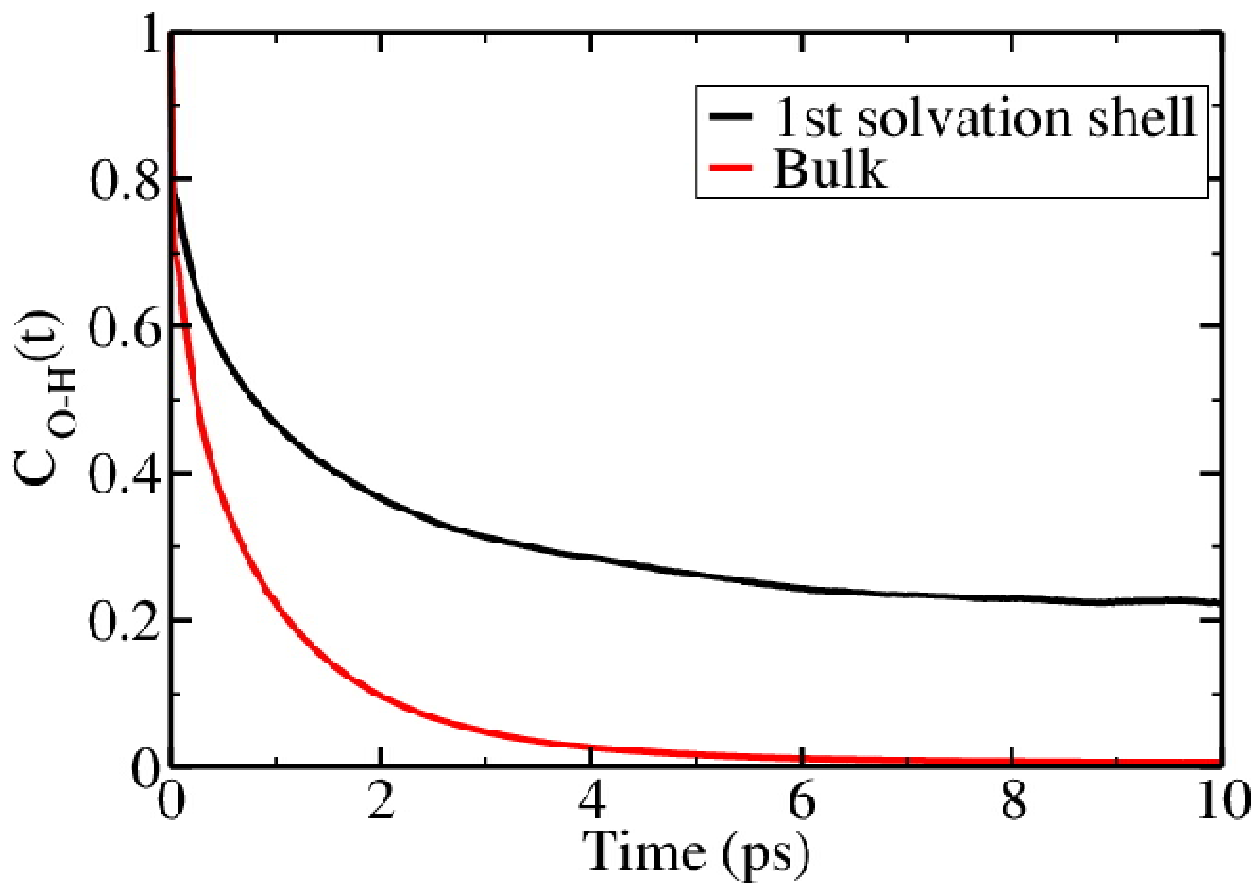

**FIG. 9. Reorientational time correlation function (TCF) of the O-H bond vector of water molecules near the cluster surface. The black curve shows the decay of TCF for the system of 16 curcumin monomers in water. The decay of TCF of bulk water is also shown in red for comparison. Extensive slow dynamics near aggregated solute with respect to bulk is noticeable. We find the same slow dynamics of water near cluster surface for all system sizes.**

We also calculated solvation time correlation function of curcumin in water. The relaxation of solvation time correlation function is demonstrated in **Fig. S4** in the supplementary material [22]. It shows a good fit to bi-exponential decay with a ultrafast one with a jump of 63% and the slow component with a jump of 36%.Avarage solvation energy of curcumin is 278.5 kJ/mol with a mean square fluctuation of energy 18 kJ/mol. Solvation dynamics of water is showing a ultrafast component which can be explained by the collective intermolecular vibrational and librational motion of water around the solute [57, 58].

## V. CONCLUSION

We have carried out the simulation of aggregation of a number of curcumin molecules in water, probably for the first time. These molecules tend to form a large cluster with a specific pattern discussed above. We monitor the pathway of aggregation that seems to evolve through a metastable configurational intermediate. The dynamical transition from segregated species to aggregated curcumin has also been described in detail by elucidating the free energy contour. We also studied the structural and orientational arrangement of surrounding water molecules and their dynamics near the cluster surface. They are rather different from bulk water showing distinct characteristics as the water molecules near a surface of a small hydrophobic solute. The aggregation property of curcumin in water is similar in nature, in certain ways, to the aggregation of amyloid beta protein (or, probably other protein molecules too) which is also completely driven by hydrophobic interaction[59]. Relative binding energy of curcumin molecules to themselves with respect to the binding energy of curcumin to beta amyloid is of great importance and therefore has far reaching consequences in curing Alzheimer disease.

The orientational preference along the system size can be explained by the energetics involved in molecular interaction of curcumin. For small system size, the entropic gain due to the anti-parallel orientation of phenyl rings out weights the energy driven parallel stack. With increasing system size, parallel orientation also predominates along with anti-parallel ones. The driving force behind the assembly is a competition between the enthalpy and entropy cost. For small system size curcumin monomers tend to form linear chain like structures with predominating anti-parallel orientation which is entopically favored, but as we increase the system size these chain like structures transform to a highly packed globuler form in which parallel orientation also emerges due to enthalpic contribution with tremendous decrease in entropy.

Hydrophobic units of curcumin do not form H-bond with water and enlarge the excluded volume in the core of the cluster from where the water molecules start to disappear. As a result it causes a drying transition which is completely due to the energetic cost as shown by Stillinger [44]. However the presence of hydrophilic groups along with the immediate hydrophobic sites in the surface together determine the orientation of water molecules that lead the aggregated structure to a sparingly soluble cluster.

The bioavailability of curcumin is related to its low solubility in water. The low solubility of curcumin monomer is related to two phenyl rings connected by a conjugated hydrocarbon chain. The solute always forms aggregates. Then how could one measure its solubility?

In order to evaluate the solvation characteristics of a molecule we often use different parallel approaches. One is Kirkwood-Buff theory. It is a statistical thermodynamic framework which provides the preferential interaction parameter from molecular distribution function of a solute with other co-solutes in a solvent mixture. In solution, for such predominantly hydrophobic molecule, the difficulty is that we always get an aggregated structure in equilibrium. While we can obtain a solute-solute distribution function with an isolated monomer that is not concentration dependent, we cannot obtain the same from the aggregated solutes. The other alternative is potential of mean force (PMF) where this approach is also not useful in the present context [60, 61]. The third alternative is to estimate the force or energy needed to bring a solute s from infinity with the other co-solute kept fixed in the solvent mixture. However, such a method is feasible for small molecule like methane. For a large molecule like curcumin, such method appears to be extremely difficult because of many possible orientations that need to be considered.

From the above discussion it is clear that we do not yet have a satisfactory even semi-quantitative method to estimate the PMF between two large predominantly hydrophobic solutes. This could be a worthwhile problem to pursue.

## ACKNOWLEDGEMENT

We thank Dr. Rakesh Sharan Singh, Dr. Rajib Biswas and Mr. Saikat Bannerjee for all the helpful discussions. This work was supported in parts by grant from DST, India. B Bagchi thanks JC Bose Fellowship (DST) for partial support of the research.